\documentclass[twocolumn,noshowpacs, preprintnumbers, amsmath,amssymb,nopreprintnumbers]{revtex4}
\usepackage{changes}
\usepackage[colorlinks,urlcolor=blue]{hyperref}
\usepackage{graphicx}
\usepackage{lineno}
\usepackage{dcolumn}
\usepackage{bm}
\usepackage{appendix}
 \usepackage{verbatim}

\usepackage{breakurl}
\usepackage{amsmath}
\usepackage{enumerate}
\makeatletter
  \let\@font@info\@gobble
  \let\@font@warning\@gobble
\makeatother

\begin{document}

\title{Statistics of a Large Number of Renewals in Equilibrium and Non-Equilibrium Renewal Processes
}

\author{Wanli Wang}
\email[contact author: ]{wanliiwang@163.com}
\affiliation{School of Mathematical Sciences, Zhejiang University of Technology, Hangzhou 310023, China}

\altaffiliation{}
\author{Stanislav Burov}

\email[contact author: ]{stasbur@gmail.com}
\affiliation{Department of Physics, Bar Ilan University, Ramat-Gan 52900, Israel}

\date{\today}

\begin{abstract}
The renewal process is a key statistical model for describing a wide range of stochastic systems in Physics. This work investigates the behavior of the probability distribution of the number of renewals in renewal processes in the short-time limit, with a focus on cases where the number of renewals is large.
We find that the specific details of the sojourn time distribution $\phi(\tau)$ in this limit can significantly modify the behavior in the large-number-of-renewals regime.
We explore both non-equilibrium and equilibrium renewal processes, deriving results for various forms of $\phi(\tau)$.
Using saddle point approximations, we analyze cases where $\phi(\tau)$ follows a power-series expansion, includes a cutoff, or exhibits non-analytic behavior near $\tau = 0$.
Additionally, we show how the short-time properties of $\phi(\tau)$ shape the decay of the number of renewals in equilibrium compared to non-equilibrium renewal processes.
The probability of the number of renewals plays a crucial role in determining rare event behaviors, such as Laplace tails.
The results obtained here are expected to help advance the development of a theoretical framework for rare events in transport processes in complex systems.

\end{abstract}

%

%


\maketitle

\section{Introduction}

Recent studies have reported the emergence of Laplace tails in probability density functions (PDFs) across various experiments. While the positional PDF of tracked particles typically exhibits Gaussian behavior for common positions, the statistics of rare events reveal a distinct pattern. In many systems, the PDF exhibits an exponential decay, often referred to as Laplace tails~\cite{hamdi2024laplace}.
These tails were observed in various systems like nanoparticle-polymer  mixtures~\cite{Hu2023Triggering}, insulin granules~\cite{yi2024distinct}, granular materials under shear~\cite{yuan2024creep}, chloroplasts in plant cells~\cite{schramma2023chloroplasts}, human cells~\cite{aaberg2021glass}, lipid bilayers~\cite{kumar2023anomalous}, spin diffusion~\cite{zu2021emergent} and many others~\cite{Kege2000Direct,Masolivera200dynamic,Weeks2000Three,Pinaki2007Universal,Wang2009Anomalous,Hapca2009Anomalous,Leptos2009Dynamics,Eisenmann2010Shear,Toyota2011Non,Skaug2013Intermittent,Xue2016Probing,Wang2017Three,Jeanneret2016Entrainment,Chechkin2017Brownian,Thomas2017Cytoplasmic,Cherstvy2019Non,Witzel2019Heterogeneities,Shin2019Anomalous,Singh2020Non,Mejia2020Tracer,Xue2020Diffusion}.
Several models like the diffusing-diffusivity, Brownian yet non-Gaussian diffusion and Fickian yet-non Gaussian diffusion~\cite{Wang2009Anomalous,Wang2012Brownian,Chubynsky2014Diffusing,Chechkin2017Brownian,hidalgo2021cusp,yamamoto2021universal,nampoothiri2021polymers} provided phenomenological routes to the observed phenomenon.
Chaudhuri, Berthier, and Kob put forward an approach that leads to the appearance of Laplace tails and is based on hop diffusion, where the particle hops/jumps between different locations/states~\cite{Pinaki2007Universal}.
It was later proven~\cite{Barkai2020Packets} that any dynamics that involve jumps and pauses between jumps, i.e., sojourn times, lead to exponential decay of the PDF~\cite{Adrian2021Large,Wang2020Large}.
Thus, Laplace tails are universal in many systems where the transport is driven by hops.

Mathematically, the exponential decay arises from the short-time behavior of the PDF of sojourn times between hops. The necessary condition for the sojourn time PDF, $\phi(\tau)$, is the existence of a power series expansion in the $\tau \to 0$ limit:
\begin{equation}\label{StasEliPlr}
\phi(\tau)\sim \sum_{j=0}^\infty C_{A+j}\tau^{A+j}
\end{equation}
where $A \geq 0$ is an integer. As long as this form of $\phi(\tau)$ holds, the probability $Q_t(N)$, representing the occurrence of $N$ events in a renewal process during time $t$, follows a universal form in the large $N/t$ limit~\cite{Barkai2020Packets,Burov2022Exponential}. This universality in $Q_t(N)$ directly leads to the universal exponential decay of the positional PDF, or Laplace tails, making the properties of $Q_t(N)$ essential for understanding this phenomenon.


In this manuscript, we focus on the statistics of the number of events, or renewals, governed by the sojourn time PDFs. While previous works~\cite{Barkai2020Packets,Burov2022Exponential} used properties of the Kummer function of the first kind, we adopt a more straightforward and broadly applicable approach: the saddle point approximation. We rederive the behavior of $Q_t(N)$ under the condition that $\phi(\tau)$ follows the form in Eq.~\eqref{StasEliPlr}, and extend the results of~\cite{Barkai2020Packets,Burov2022Exponential} to cases where $\phi(\tau)$ deviates from this form. This includes scenarios where the leading term in the power series expansion has a non-integer or negative exponent.
Additionally, we analyze $Q_t(N)$ for cases where $\phi(\tau)$ is non-analytic near $\tau=0$ or strictly zero at this limit. We demonstrate the validity of our approach using specific PDFs, such as the Lévy and Mittag-Leffler distributions. Notably, we show that the large-$N$ decay of $Q_t(N)$ differs between the Lévy and Mittag-Leffler cases, in contrast to the behavior of typical fluctuations~\cite{Tian2020Continuous,Liu2022Strong,Zhou2022Generalized,Liu2023Levy}.


We consider two types of renewal processes: the non-equilibrium renewal process, which starts at time $t=0$ (also referred to as the regular renewal process), and the equilibrium renewal process, where observations begin after the system has been running for an extended period.
While it is often assumed for computational convenience that the process starts at the same time as the observation, this does not reflect many experimental situations. Frequently, the system evolves for some time before observations or tracking begins, allowing it to reach steady-state or equilibrium.
Therefore, understanding the probability of the number of events in equilibrium renewal processes becomes crucial. We examine this probability for various forms of $\psi(\tau)$ in the large $N$ limit.



The manuscript is structured as follows: Sec.~\ref{model} introduces the normal renewal process and derives the general form of $Q_t(N)$ in Laplace space.
In Sec.~\ref{model:power}, we derive the explicit form of $Q_t(N)$ for a generalized power-series expansion of $\phi(\tau)$, tested using the Mittag-Leffler distribution.
Sec.~\ref{model:cutoff} addresses the case of vanishing $\phi(\tau)$, and Sec.~\ref{model:nonanalytic} covers non-analytic $\phi(\tau)$ in the $\tau \to 0$ limit.
Sec.~\ref{AT19APP5} examines neglected terms in previous sections using the L{\'e}vy distribution.
Sec.~\ref{EquilRenewal} derives $Q_t(N)$ for the equilibrium renewal process with both a generalized power-law expansion (Sec.~\ref{equillb:power}) and a cut-off (Sec.~\ref{equillib:cutoff}).
The final section provides discussions and conclusions.




\section{Ordinary Renewal Process}\label{model}
The renewal process describes times when events, like the emission of photons or the birth of a new bacteria, occur~\cite{Cox1977Theory,Godreche2001Statistics,Wang2018Renewal}.
The ordinary renewal process, also known as the non-equilibrium renewal process, starts at time $t=0$, and the first event occurs at time $t_1$, the second event at $t=t_2$ and the $n$th event at $t=t_N$.
While the time of occurrence of the $N$th event depends on the time of occurrence of the $(N-1)$th event, the time intervals $\tau_N=t_N-t_{N-1}$ are assumed to be positive, independent, and identically distributed (IID) random variables. $\phi(\tau)$ is the PDF of the time intervals $\tau_N$.
The process is terminated at some general time $t$, and the key question is the number of events that occurred up to $t$, i.e., $N_t$. Since all the $\tau_N$ are random, the value of $N_t$ is also a random variable.
In the following, we drop the subscript describing the time dependence of $N_t$ and assign $Q_t(N)$ as the probability of observing $N$ events in the time interval $(0,t)$. This probability is determined by the fact that $t_N\leq t<t_{N+1}$, and therefore $Q_t(N)=\langle I(t_N\leq t<t_{N+1})\rangle$, where $I(\dots)$ is the indicator function and $\langle \dots\rangle$ is averaging over all possible realizations of the process.
Since $\int_0^\infty e^{-st} I(t_N\leq t<t_{N+1})\,dt=\frac{1}{s}\left(e^{-st_N}-e^{-st_{N+1}}\right)$, and $t_N=\tau_1+\tau_2+\cdots+\tau_N$, the Laplace transform $\widehat{Q}_s(N)=\int_0^\infty e^{-st} Q_t(N)\,dt$ reads~\cite{Godreche2001Statistics}
\begin{equation}\label{AT19SEC2EQ1202}
 \widehat{Q}_s(N)=\frac{1-\widehat{\phi}(s)}{s}\widehat{\phi}^N(s),
\end{equation}
where $\widehat{\phi}(s)=\mathcal{L}[\phi(\tau)]=\int_0^\infty e^{-s\tau}\phi(\tau)\,d\tau$ is the Laplace transform of  $\phi(\tau)$ and $\mathcal{L}^{-1}[\widehat{\phi}(s)]$ denotes the corresponding inverse Laplace transform operator for $\widehat{\phi}(s)$.
Eq.~\eqref{AT19SEC2EQ1202} allows to obtain the behavior of $Q_t(N)$ in various limits of $t$ and $N$.
The limit of a long time, i.e., $t\to\infty$, was explored for various sojourn time PDFs, as discussed in~\cite{Godreche2001Statistics,Schulz2014Aging,Wang2018Renewal}. However, as was originally noted in~\cite{Barkai2020Packets}, it is the limit of large $N$, and specifically large $N/t$, that is of crucial importance for the description of rare events and Laplace tails. Therefore, we focus on the short observation time $t$ and large $N$.

In this short $t$ and large $N$ limit, only the short $\tau$ properties of $\phi(\tau)$ are important. It is irrelevant what is the probability of a very long time interval since it can not occur during time $t$, given that $t$ is short.
For $\tau\to 0$ we make a distinction between two cases: (a) $\lim_{\tau\to 0}\phi(\tau)$ is finite (including $0$) and (b)  $\lim_{\tau\to 0}\phi(\tau)$ diverges as $\tau^{\alpha-1}$ while $0<\alpha<1$, any stronger divergence is impossible due to the normalization condition for $\phi(\tau)$.
In Laplace space, for both (a) and (b), $\lim_{s\to\infty}{\widehat\phi}(s)\to 0$.
Indeed, for (a), according to the initial value theorem
\begin{equation}
\lim_{\tau\to 0^{+}}\phi(\tau)=\lim_{s\to \infty}s\widehat{\phi}(s)
\end{equation}
and therefore ${\widehat\phi}(s)\to 0$ in the large $s$ limit. For (b), the Tauberian theorem~\cite{Feller1971introduction} states that in the limit $s\to\infty$, ${\widehat \phi}(s)\propto s^{\alpha-2}$, and therefore ${\widehat\phi}(s)\to 0$ in the large $s$ limit.
Due to this convergence of ${\widehat \phi}(s)$ to $0$,
from Eq.~\eqref{AT19SEC2EQ1202}  it follows that in the $s\to \infty$ limit the leading behavior of ${\widehat Q}_s(N)$ is provided by ${\widehat Q}_s(N)\sim{\widehat \phi}(s)^N/s$ (see the discussion in Sec.\ref{AT19APP5}).
According to the Tauberian theorem, for $t\to 0$, $Q_t(N)$ is determined by the $s\to\infty$ behavior of ${\widehat Q}_s(N)$ and therefore in the following we use
\begin{equation}\label{AT19SECassa}
\widehat{Q}_s(N)\sim\frac{\exp(N\ln(\widehat{\phi}(s)))}{s}.
\end{equation}
Equation~\eqref{AT19SECassa} is utilized below to explore various possible behaviours of $Q_t(N)$.
In contrast to the approach outlined in \cite{Barkai2020Packets}, Eq.~\eqref{AT19SECassa} offers a considerably more straightforward and simplified way of obtaining the appropriate functional form of $Q_t(N)$.
Below, we consider four general cases of $\phi(\tau)$ and use Eq.~\eqref{AT19SECassa} to derive the functional form of $Q_t(N)$ in the large $N/t$ limit.

\subsection{The case of power-series expansion of $\phi(\tau)$} \label{model:power}

In this section, we assume a generalization of the scenario provided by Eq.~\eqref{StasEliPlr}. Namely,
\begin{equation}\label{phitausMALL}
\phi(\tau)=C_\alpha \tau^\alpha+C_\beta \tau^\beta+\cdots,~~~~(\tau\to 0),
\end{equation}
where $\alpha>-1$ and $\beta>\alpha$.
Notice that for Eq.~\eqref{phitausMALL} $\alpha$ and $\beta$ don't have to attain integer values.
 When $\alpha$ is in the range  $-1<\alpha<0$, $\phi(\tau)$ diverges as $\tau\to 0$.
 Conversely, when $\alpha> 0$, $\phi(\tau)$ converges to $0$ and when $\alpha=0$ the distribution $\phi(\tau)$ converges to a constant value, i.e., $\phi(\tau)\to C_\alpha$.
As mentioned above, we are interested in the behavior of small observation time $t$, which leads to $s\to\infty$.
Using the Laplace pair $\tau^\alpha\to s^{-\alpha-1} \Gamma (\alpha+1)$, the Laplace transform of Eq.~\eqref{phitausMALL} reads
\begin{equation}\label{sdi101}
\widehat{\phi}(s)\sim C_\alpha\frac{ \Gamma(\alpha+1)}{s^{\alpha+1}}+C_\beta\frac{ \Gamma(\beta+1)}{s^{\beta+1}},
\end{equation}
when $s\to \infty$.
Eq.~\eqref{AT19SECassa}, then yields
\begin{equation}\label{Seq101}
\begin{split}
\widehat{Q}_s(N)&\sim \frac{1}{s}\left(\frac{C_\alpha \Gamma(\alpha+1)}{s^{\alpha+1}}+\frac{C_\beta \Gamma(\beta+1)}{s^{\beta+1}}\right)^N.
\end{split}
\end{equation}
Since $s\to \infty$ and $N$ is large, Eq.~\eqref{Seq101} reduces to
\begin{equation}\label{1Seq102}
\widehat{Q}_s(N)\sim \frac{(C_\alpha \Gamma(\alpha+1))^N}{s^{(\alpha+1)N+1}}\exp\left(\frac{C_\beta \Gamma(\beta+1)}{C_\alpha \Gamma(\alpha+1)s^{\beta-\alpha}}\right).
\end{equation}
Then, according to the definition of the inverse Laplace transform,  Eq.~\eqref{1Seq102} leads to
\begin{widetext}
\begin{equation}\label{eshsh210}
Q_t(N)\sim\frac{(C_\alpha \Gamma(\alpha+1))^N}{2\pi i}\int_{c-i\infty}^{c+i\infty}\exp\left(\underbrace{st-(1+(\alpha+1)N)\ln(s)+N\frac{C_\beta \Gamma(\beta+1)}{C_\alpha \Gamma(\alpha+1)s^{\beta-\alpha}}}_{\chi(s)}\right)ds.
\end{equation}
\end{widetext}
Notice that here $t$ is kept constant, and $N$ is large.
The inverse Laplace transform is calculated using the saddle point method
\begin{equation}\label{sssad}
\frac{1}{2\pi i}\int_{c-i\infty}^{c+i\infty}\exp(N\chi(s))ds\sim \frac{\exp(N\chi(s^{*}))}{\sqrt{2\pi N\chi^{''}(s^{*})}},
\end{equation}
where $s^*$ satisfies $\chi^{'}(s)|_{s=s^*}=0$.
Specifically for Eq.~\eqref{eshsh210} we obtain that $s^*$ satisfies
\begin{equation}\label{ssssa}
t=\frac{N}{s^*}\left[(\alpha+1)+\frac{1}{N}+\frac{\alpha-\beta}{(s^*)^{\alpha-\beta}}\frac{C_\beta\Gamma(1+\beta)}{C_\alpha\Gamma(1+\alpha)}\right],
\end{equation}
and since $N$ is large and $t$ is kept constant, $s^*$ has to be large as well.
In this limit of large $s^*$ we utilize the condition $\alpha<\beta$ and Eq.~\eqref{ssssa} yields
\begin{equation}\label{esjdjj2}
s^*\sim\frac{(\alpha+1)N+1}{t}.
\end{equation}
Notice that the condition of large $s^*$ justifies the discussed $s\to\infty$ limit for ${\widehat Q}_s(N)$ in Eq.~\eqref{AT19SECassa}.

Utilizing Eqs.~\eqref{eshsh210} and \eqref{esjdjj2}, we obtain in the large $N/t$ limit
\begin{widetext}
\begin{equation}\label{1seq201}
Q_t(N)\sim \frac{\exp(N\ln[C_\alpha \Gamma(1+\alpha)])}{\sqrt{2\pi t^2/(1+(1+\alpha)N)}}\exp\left(-((\alpha+1)N+1)\ln\left(\frac{(\alpha+1)N+1}{te}\right)+N^{1+\alpha-\beta}(1+\alpha)^{\alpha-\beta}\frac{C_\beta\Gamma(1+\beta)}{C_\alpha \Gamma(1+\alpha)}t^{\beta-\alpha}\right)
\end{equation}
\end{widetext}
and by using Stirling's approximation for the $\Gamma$ function (~$\Gamma(z+1)\approx \sqrt{2\pi z}z^z\exp(-z)$~), we finally derive our main result of this subsection, namely
\begin{equation}\label{shssheq101}
\begin{split}
Q_t(N)&\underset{N/t\to\infty}{\sim} \frac{([C_\alpha\Gamma(1+\alpha)]^{1/(1+\alpha)}t)^{N(1+\alpha)}}{\Gamma((\alpha+1)N+1)}\\
&\times\exp\left(\frac{N^{1+\alpha-\beta}}{(1+\alpha)^{\beta-\alpha}}\frac{C_\beta\Gamma(1+\beta)}{C_\alpha \Gamma(1+\alpha)}t^{\beta-\alpha}\right).
\end{split}
\end{equation}
The form for $Q_t(N)$ as it appears in Eq.~\eqref{shssheq101} is a generalization of the original result obtained in~\cite{Barkai2020Packets} and is applicable to a much wider class of problems.
For the special case when $\beta-\alpha=1$,
Eq.~\eqref{shssheq101} reduces to
\begin{equation}\label{StasExp}
Q_t(N)
\underset{N/t\to\infty}{\sim} \frac{([C_\alpha\Gamma(1+\alpha)]^{\frac{1}{1+\alpha}}t)^{N(1+\alpha)}}{\Gamma((\alpha+1)N+1)}\exp\left(\frac{C_{\alpha+1}}{C_\alpha }t\right),
\end{equation}
which is identical to the result obtained in \cite{Barkai2020Packets}.
Another important limit holds when $1+\alpha-\beta<0$ and the exponent on the right hand side of Eq.~\eqref{shssheq101} can be ignored, namely
\begin{equation}\label{equ3ow}
Q_t(N)\underset{N/t\to\infty}{\sim}  \frac{[C_\alpha\Gamma(\alpha+1)]^Nt^{N(\alpha+1)}}{\Gamma(N(\alpha+1)+1)}
.
\end{equation}

\begin{figure}[t]
 \centering
 \includegraphics[width=0.5\textwidth]{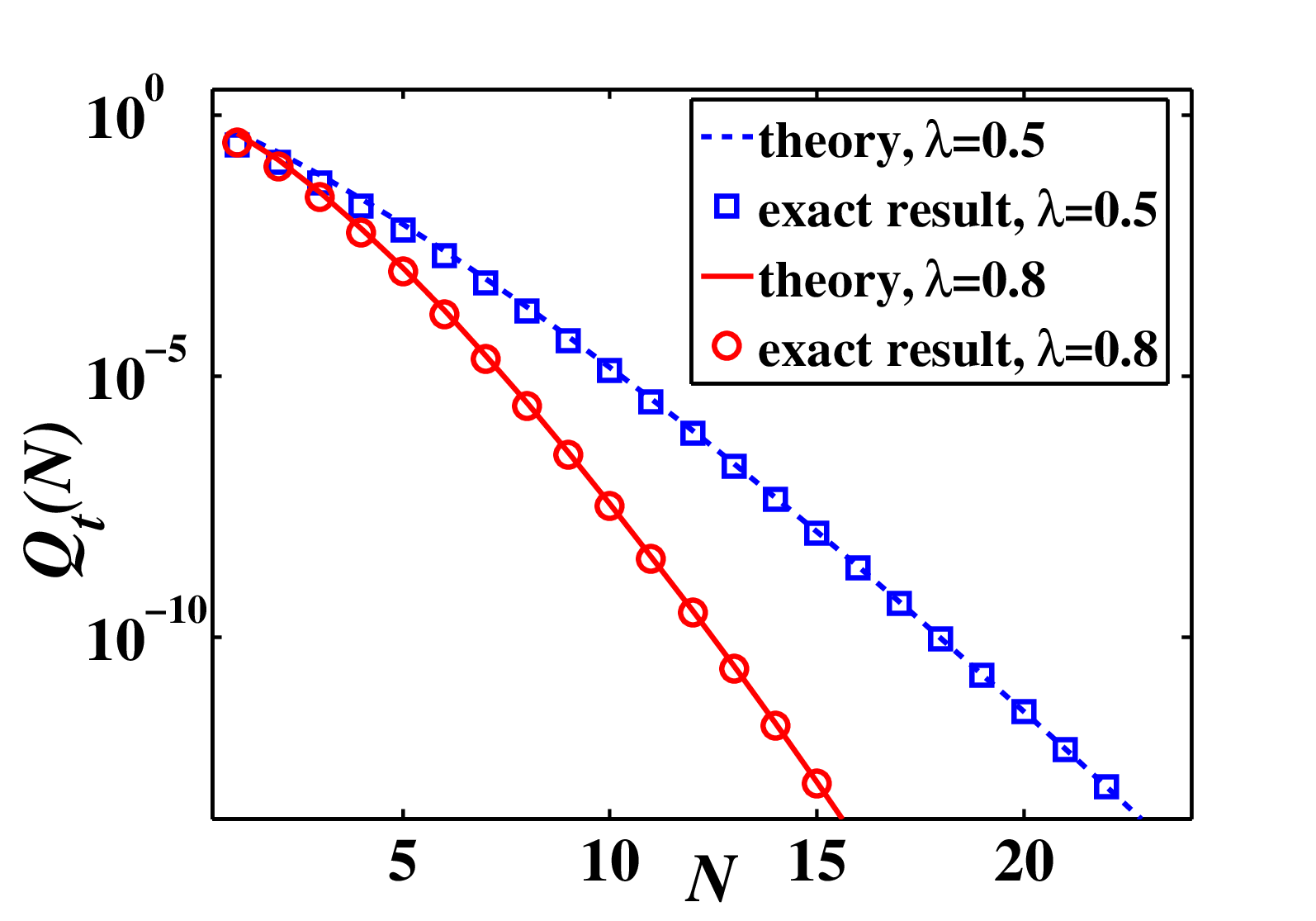}\\
 \caption{Appearance of the nearly exponential tails, with sojourn times drawn from the Mittag-Leffler distribution given in Eq.~\eqref{3SE101}. The symbols represent the exact result from Eq.~\eqref{MLFeXACT}, while the corresponding theoretical prediction, obtained from Eq.~\eqref{1seq201}, is plotted as lines. Here, we choose $t=0.5$, $\alpha=\lambda-1$, $\beta=2\lambda-1$, $C_\alpha=1/\Gamma(\lambda)$, and $C_\beta=-1/\Gamma(2\lambda)$.
}\label{MLFtHalf}
\end{figure}

As a verification of our results in this subsection, we consider a special case when $\phi(\tau)$ is the Mittag-Leffler distribution \cite{Haubold2011Mittag}, which is relevant for many systems where anomalous transport takes place~\cite{Metzler2000random,burovbarkai2008PhysRevE}.
The Mittag-Leffler distribution has the form
\begin{equation}\label{3SE101}
\phi(\tau)=\tau^{\lambda-1}E_{\lambda}(-\tau^\lambda)
\end{equation}
with $0<\lambda<1$ and $E_{\lambda}(y)=\sum_{j=0}^\infty y^j/\Gamma(\lambda j+\lambda)$.
It is well-known that the far tail of $\phi(\tau)$ decays as a power-law $\tau^{-\lambda-1}$ \cite{Podlubny1999Fractional}.
For small $\tau$,
\begin{equation}
\phi(\tau)\sim \frac{\tau^{\lambda-1}}{\Gamma(\lambda)}-\frac{\tau^{2\lambda-1}}{\Gamma(2\lambda)},
\end{equation}
where we used the relation $E_{\lambda}(z)\sim 1/\Gamma(\lambda)+z/\Gamma(2\lambda)$ for $z\to 0$ \cite{Cahoy2013Estimation}.
This indicates that $\phi(\tau)$ diverges in the limit when $\tau\to 0$.
The Mittag-Leffler distribution doesn't satisfy Eq.~\eqref{StasEliPlr}, except when $\lambda$ is a positive integer and the form of $Q_t(N)$, as provided in \cite{Barkai2020Packets}, is not valid.
We use the Mittag-Leffler distribution to verify the applicability of Eq.~\eqref{shssheq101}.
The form of $\phi(\tau)$ in Eq.~\eqref{3SE101} allows an exact representation of $Q_t(N)$ as we now show.
The Laplace transform of Eq.~\eqref{3SE101} yields
\begin{equation}\label{3SE102}
\widehat{\phi}(s)=\frac{1}{1+s^\lambda},
\end{equation}
then according to Eq.~\eqref{AT19SEC2EQ1202}
\begin{equation}\label{3SE103}
\widehat{Q}_s(N)=\frac{s^{\lambda-1}}{(1+s^\lambda)^{N+1}}
\end{equation}
and the
 inverse Laplace
 yields the exact expression
\begin{equation}\label{MLFeXACT}
Q_t(N)=t^{\lambda N}E_{\lambda,1+\lambda N}^{(N+1)}(-t^\lambda),
\end{equation}
where $
E^{(n)}_{\lambda,\xi}(z)=\sum_{j=0}^\infty \Gamma(n+j) z^j\Big/{j!\Gamma(n)\Gamma(\xi+\lambda j)}$, is the $n$th order derivative of Mittag-Leffler function $E_{\lambda,\xi}(z)$.
In Fig.~\ref{MLFtHalf}, the comparison between the exact behavior of the $Q_t(N)$ for the case of Mittag-Leffler distribution (Eq.~\eqref{MLFeXACT}) and the asymptotic relation, provided by Eq.~\eqref{1seq201}, shows an excellent agreement.
In the next subsection, we extend our exploration of the form of $Q_t(N)$ to the case when $\phi(\tau)$ is identically zero in the vicinity of $\tau=0$.

\subsection{The case of $\phi(\tau)$ with a cutoff}\label{model:cutoff}

\begin{figure}[htb]
 \centering
 \includegraphics[width=0.5\textwidth]{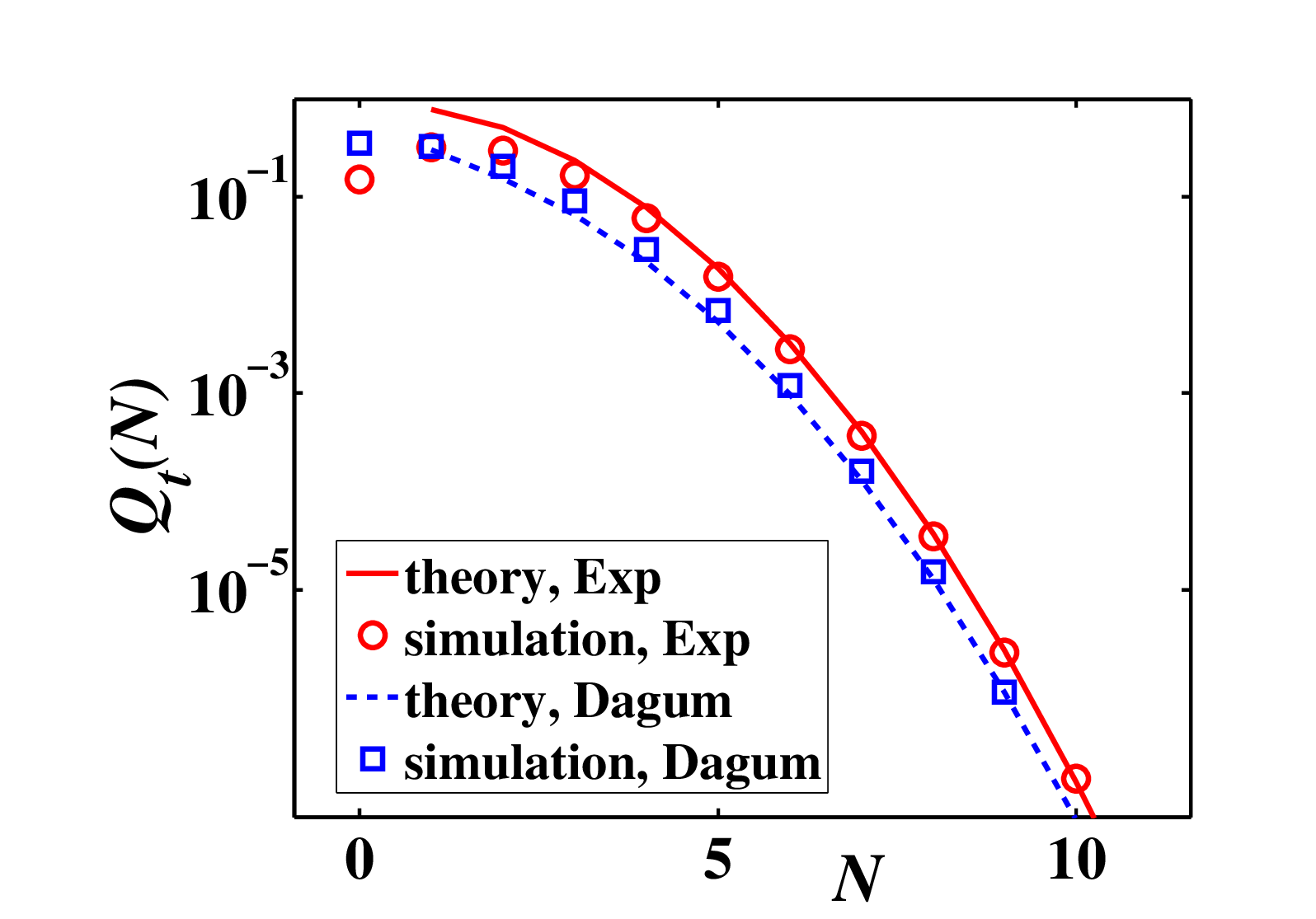}\\
 \caption{Behavior of $Q_t(N)$ versus $N$ as obtained using a cutoff for $\phi(\tau)$ provided in Eq.~\eqref{cutoff101} with $\tau_0=0.1$. The lines are the plot of Eq.~\eqref{sjdhs101} versus $N$ for different sojourn time PDFs, i.e., exponential distribution $\phi(\tau)=\exp(-(t-\tau_0))\Theta(t-\tau_0)$ and Dagum distribution $\phi(\tau)=1/(1+\tau-\tau_0)^2\Theta(\tau-\tau_0)$. The symbols describe the simulation results with $t=2$.
}\label{QtNcutOff}
\end{figure}

The power-series expansion, as it appears in Eq.~\eqref{phitausMALL}, assumes a non-zero probability of observing $\tau>0$.
This subsection is dedicated to the case when the probability of observing $\tau<\tau_0$ is zero. Namely, we introduce a cutoff for $\phi(\tau)$.
The power-series expansion of $\phi(\tau)$ holds for any $\tau>\tau_0>0$, i.e.,
\begin{equation}\label{cutoff101}
\phi(\tau)\sim \left(C_\alpha(\tau-\tau_0)^\alpha+C_\beta(\tau-\tau_0)^\beta+\cdots \right)\Theta(\tau-\tau_0),
\end{equation}
where $\Theta(z)$ is the Heaviside  function:   $\Theta(z)=0$ for $z<0$ and $\Theta(z)=1$ for $z>0$.
The constants $C_\alpha$, $C_\beta$, $\alpha$, and $\beta$ are determined by the exact form of $\phi(\tau)$.
The presence of the Heaviside function introduces the cutoff at $\tau_0$.
As for the case of Eq.~\eqref{phitausMALL}, we assume that $-1<\alpha<\beta$.
Since
$
\int_0^\infty \exp(-s\tau)(\tau-\tau_0)^\alpha \Theta(\tau-\tau_0) d\tau={\exp(-s\tau_0)\Gamma(1+\alpha)}\big/{s^{1+\alpha}}$,  Eq.~\eqref{AT19SECassa} for the case of $\phi(\tau)$ with a cutoff yields
\begin{equation}
\begin{split}
  \widehat{Q}_s(N)\sim  & \frac{(C_\alpha \Gamma(1+\alpha))^N}{s^{(1+\alpha)N+1}} \\
  &\times\exp\left(-s\tau_0N+\frac{NC_\beta\Gamma(1+\beta)}{C_\alpha\Gamma(1+\alpha)s^{\beta-\alpha}}\right).
\end{split}
\end{equation}
The form of $Q_t(N)$ in the large $N/t$ limit is again found using the saddle point approximation, where
\begin{equation}
    \label{eq:sstarforcutoff}
s^*=\frac{(\alpha+1)N+1}{t-\tau_0N}
\end{equation}
that leads to
\begin{equation}\label{sjdhs101}
\begin{split}
Q_t(N)&\sim \frac{[(C_\alpha \Gamma(1+\alpha))^{\frac{1}{1+\alpha}}(t-N\tau_0)]^{N(1+\alpha)}}{\Gamma((\alpha+1)N+1)}\\
&~~~\times\exp\Big(\frac{C_\beta\Gamma(1+\beta)}{C_\alpha\Gamma(1+\alpha)}\frac{N(t-\tau_0 N)^{\beta-\alpha}}{((\alpha+1)N+1)^{\beta-\alpha}}\Big)\\
&~~~\times \Theta(t-\tau_0N).
\end{split}
\end{equation}
Notice that the presence of the $\Theta$ function for $\phi(\tau)$ leads to the presence of $\Theta(t-\tau_0 N)$ in $Q_t(N)$. It means that for any $N>t/\tau_0$, $Q_t(N)$ is identically $0$.
This is easily understood by noting that the minimal possible $\tau$ is $\tau_0$; therefore, for an observation time $t$, the maximal number of events is $t/\tau_0$.
The applicability of  Eq.~\eqref{sjdhs101} holds when $s^*\to\infty$, i.e.,
$$s^{*}=\frac{1+(\alpha+1)N}{t-\tau_0N}\simeq \frac{\alpha+1}{\tau_0}\frac{ \tau_0N/t}{1-\tau_0N/t}\to \infty.$$
Since $y/(1-y)$ is an increasing function of $y=\tau_0N/t< 1$, as we increase $N$ and approach the $\tau_0 N/t \to 1$ limit, Eq.~\eqref{sjdhs101} will hold.
In Fig.~\ref{QtNcutOff},
we present an excellent agreement between simulation results and the theoretical form in  Eq.~\eqref{sjdhs101}.

\subsection{Non-Analytic form of $\phi(\tau)$ in the vicinity of $\tau\to 0$} \label{model:nonanalytic}

\begin{figure}[htb]
 \centering
 \includegraphics[width=0.5\textwidth]{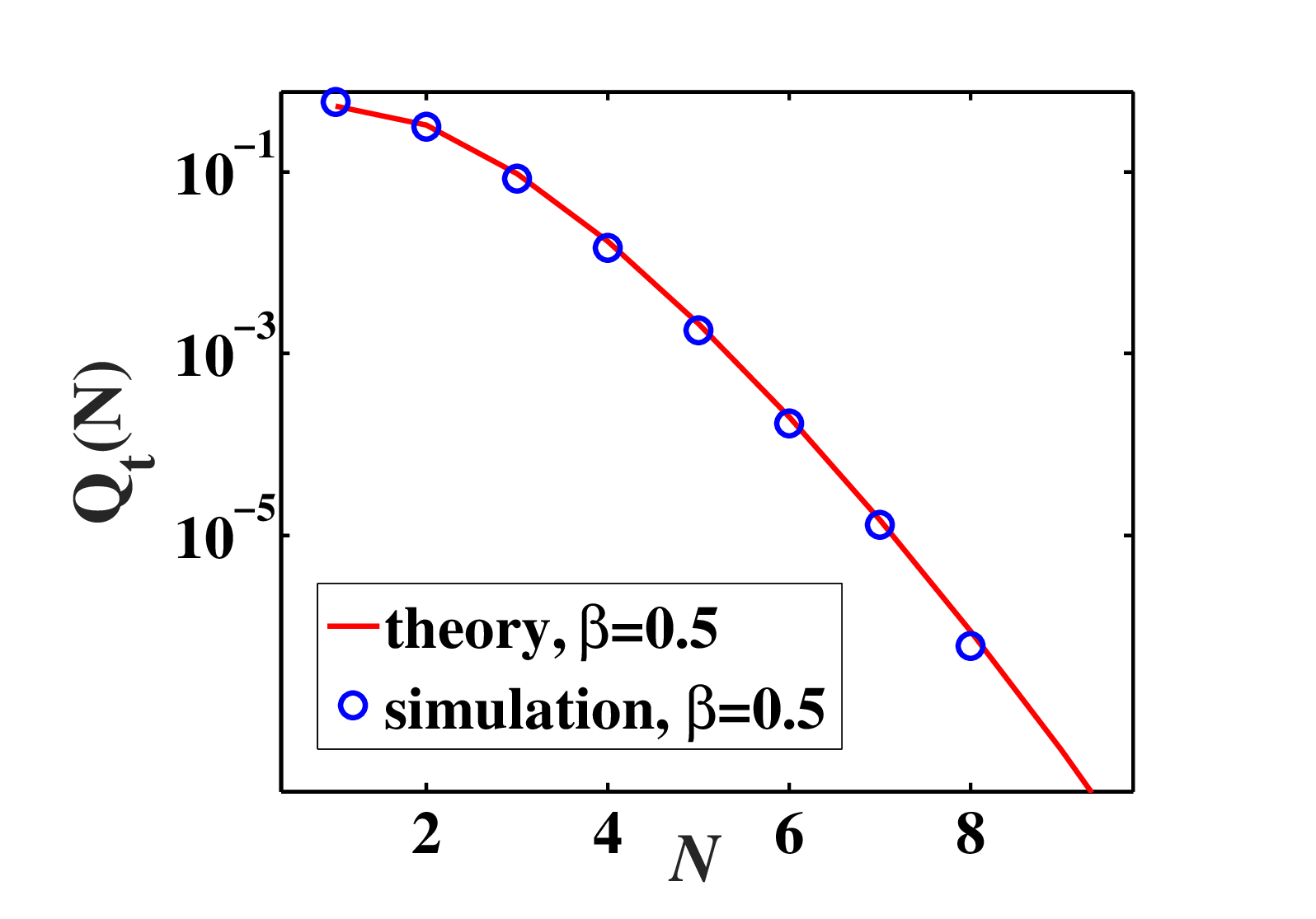}\\
    \caption{The compressed exponential decay of $Q_t(N)$ for sojourn times following Eq.~\eqref{12so101}. The solid line is the theory Eq.~\eqref{eqssa107}, and the corresponding symbols are simulations obtained from the realizations of particles. In our simulations, we use $\phi(\tau)=c\exp(-\mu/\tau^\beta)\Theta(0<\tau\leq2)$, where $c=1/\int_0^2\exp(-\mu/\tau^\beta)d\tau$.
The parameters are $t=2$, $\mu=1/4$, $\beta=0.5$, and $c\simeq1.462$.
}\label{QtNStretch}
\end{figure}

In this subsection, we treat the case when $\phi(\tau)$ is not strictly $0$ near $\tau\to 0$, but it can't be presented in terms of power series expansion in this limit.
Specifically, we are focused on the case when the approach of $\phi(\tau)$ to $0$ is extremely fast, of the form
\begin{equation}\label{12so101}
\phi(\tau)= c\exp(-\mu/\tau^\beta),
\end{equation}
where $\beta$, $\mu$, and $c$ are some positive constants.
The Laplace transform of $\phi(\tau)$ in the limit of $s\to\infty$ is achieved by the saddle-point approximation where only the small $\tau$ properties are important.
The saddle point of the integral $\int_0^\infty e^{-s\tau}\phi(\tau)\,d\tau$ is achieved for $\tau^*=(\mu\beta/s)^{1/(1+\beta)}$ and in the limit of $s\to \infty$, ${\widehat\phi}(s)$ is
\begin{equation}\label{12so102}
\begin{split}
\widehat{\phi}(s)& \sim\frac{\exp\left(-\frac{1+\beta}{\beta}(\mu\beta)^{\frac{1}{1+\beta}}s^{\frac{\beta}{1+\beta}}-\frac{(2+\beta)\ln(s)}{2(1+\beta)} \right)}{\sqrt{\mu\beta(\beta+1)(\mu\beta)^{-\frac{\beta+2}{\beta+1}}/(2\pi c^2)}}.
\end{split}
\end{equation}
Next, we substitute ${\widehat\phi}(s)$ into Eq.~\eqref{AT19SECassa} and use the saddle point approximation, similar to the two previous cases of $\phi(\tau)$ discussed above.
The saddle point occurs at $s^*$ that satisfies
\begin{equation}\label{eqssa101}
t-N (\beta  \mu )^{\frac{1}{\beta +1}} {s^*}^{-\frac{1}{\beta +1}}-\frac{(\beta +2) N}{2 (\beta +1) s^*}=0.
\end{equation}
For fixed $t$ and large $N$
the asymptotic solution reads
$s^*\sim \beta  \mu  \left({N}\big/{t}\right)^{\beta +1}$. Therefore, we seek a solution of the form
$s^{*}=\mu\beta(N/t)^{1+\beta}+\gamma N^g$, where $\gamma$ and $g$ are undetermined coefficients.
Substituting  $s^{*}$ into  Eq.~\eqref{eqssa101} and using  Taylor expansion, we obtain that
$
t\gamma N^g-N^g\gamma t\frac{\beta}{1+\beta}-N\frac{2+\beta}{2(1+\beta)}=0
$, and comparison of appropriate coefficients yields
$ t\gamma -\gamma t\frac{\beta}{1+\beta}-\frac{2+\beta}{2(1+\beta)}=0$ and $g=1$.
Thus, $s^*$ in Eq.~\eqref{eqssa101} follows
\begin{equation}\label{sssaas}
s^*\sim \beta  \mu  \left(\frac{N}{t}\right)^{\beta +1}+\frac{\beta+2}{2}\frac{N}{t}.
\end{equation}
Finally, we obtain that $Q_t(N)$ in the limit of large $N/t$ is
\begin{equation}\label{eqssa107}
\begin{split}
 Q_t(N) & \sim \frac{d^N}{\sqrt{2\pi |\chi^{''}(s^*)|}}\\
 &~~~\times\exp\Big(-N\frac{1+\beta}{\beta}(\mu\beta)^{\frac{1}{1+\beta}}(s^*)^{\frac{\beta}{1+\beta}}\\
 &~~~+s^*t-N\frac{(\beta+2)\ln(s^*)}{2(1+\beta)}\Big),
\end{split}
\end{equation}
where $d=c\sqrt{2\pi(\mu\beta)^{(\beta+2)/(\beta+1)/(\mu\beta(\beta+1))}}$ and  $\chi^{''}(s)=N \left(\beta +2 (\beta  \mu )^{\frac{1}{\beta +1}} s^{\frac{\beta }{\beta +1}}+2\right)/(2 (\beta +1) s^2)$,  while the value of $s^{*}$ is provided in Eq.~\eqref{sssaas}.
For large $N$, Eq.~\eqref{eqssa107} reduces to
\begin{equation}
\label{eq:nonalyticLarge}
\begin{split}
Q_t(N)&\sim \frac{d^N}{\sqrt{2\pi}}\exp\Big(\mu\beta N^{1+\beta}t^{-\beta}-N^{1+\beta}\mu\frac{1+\beta}{t^{\beta}}\\
&~~~~~~~~~-\frac{N(\beta+2)}{(2(1+\beta))}\ln\left(\frac{\mu\beta N^{1+\beta}}{t^{1+\beta}}\right)\Big)\\
&\sim \frac{d^N}{\sqrt{2\pi}}\exp\left(-N \mu \left[\frac{N}{t}\right]^\beta\right),
\end{split}
\end{equation}
showing compressed/squeezed exponential decay $\exp(-N^{\nu})$ \cite{Kohlrausch1854Theorie,Wuttke2012Laplace,Falus2006Crossover,Defaveri2024Stretched}.
In Fig.~\ref{QtNStretch}, an excellent agreement between the theory (Eq.~\eqref{eqssa107} ) and numerical simulation is presented.
The non-analytical properties of $\phi(\tau)$ at $\tau\to 0$ limit are also present in the $t\to 0$ limit of $Q_t(N)$ in the form of $\exp(-N^{1+\beta}\big/t^\beta)$.

\subsection{The one-sided L{\'e}vy distribution}\label{AT19APP5}

We have analyzed the behavior of $Q_t(N)$ for three cases of waiting times PDFs based on Eq.~\eqref{AT19SECassa}.
The presented approach is based on the assumption that for a large number of renewals $N$ and sufficiently short times $t$,  the distribution $Q_t(N)$ is determined by Eq.~\eqref{AT19SECassa} and the term $\widehat{\phi}(s)^{N+1}/s$ in Eq.~\eqref{AT19SEC2EQ1202}, can be disregarded.
This subsection describes why $\widehat{\phi}(s)^{N+1}/s$ is negligible when compared to $\widehat{\phi}(s)^{N}/s$.
As a case study, we consider the well-known one-sided L{\'e}vy distribution $\ell_\alpha (\tau)$ for $\phi(\tau)$ \cite{Zolotarev1986One}.
While the form of $\ell_\alpha(\tau)$ in $t$ space is cumbersome, in Laplace space $\ell_\alpha(\tau)$\cite{Schneider1986Stable} obeys
\begin{equation}\label{OneLevy}
\begin{split}
\int_0^\infty\exp(-s\tau)\ell_\alpha(\tau)d\tau=\exp(-s^\alpha)
\end{split}
\end{equation}
with $0<\alpha<1$.
For $\tau\to\infty$,  $\ell_{\alpha}(\tau)\sim\tau^{-\alpha-1}$ yielding a similar behavior to the power-law decay of the Mittag-Leffler distribution discussed above.
When $\tau\to 0$, $\ell_\alpha(\tau)$ follows \cite{Schneider1986Stable}
\begin{equation}\label{smallLevyExpansion}
\phi(\tau)\sim \sqrt{\frac{\alpha^{\frac{1}{1-\alpha}}}{2\pi(1-\alpha)}} \tau^{-\frac{2-\alpha}{2(1-\alpha)}}\exp\left(-[1-\alpha]\left(\alpha\tau\right)^{-\frac{\alpha}{1-\alpha}}\right).
\end{equation}
The properties of $\phi(\tau)$ in Eq.~\eqref{smallLevyExpansion} are determined by the extremely fast decaying term $\exp(-(1-\alpha)\alpha^{\alpha/(1-\alpha)}\tau^{-\alpha/(1-\alpha)})$.

Notice that according to Eq.~\eqref{OneLevy} we have $\widehat{\phi}(s)=\exp(-s^\alpha)$. Based on Eq.~\eqref{AT19SEC2EQ1202}
\begin{equation}\label{saddleMehodsQtn}
Q_t(N)= Q[t,\alpha,N]-Q[t,\alpha,N+1],
\end{equation}
where $Q[t,\alpha,N]={\cal L}^{-1}\left\{\frac{1}{s} \exp\left({-N s^\alpha}\right)\right\}$ and was utilized as approximation for $Q_t(N)$ (see Eq.~\eqref{AT19SECassa}).
The advantage of presenting $Q_t(N)$ using Eq.~\eqref{saddleMehodsQtn} is that it allows to examine the effect of the neglected term in Eq.~\eqref{AT19SEC2EQ1202}, i.e., $\widehat{\phi}^{N+1}(s)/s$. Namely, $Q_t(N)=Q[t,\alpha,N]\left(1-Q[t,\alpha,N+1]/Q[t,\alpha,N]\right)$, and the ratio $Q[t,\alpha,N+1]/Q[t,\alpha,N]$ displays the neglected part which is not a part of the approximation in Eq.~\eqref{AT19SECassa}.

\begin{figure}[htb]
\centering
\includegraphics[width=1.0\linewidth]{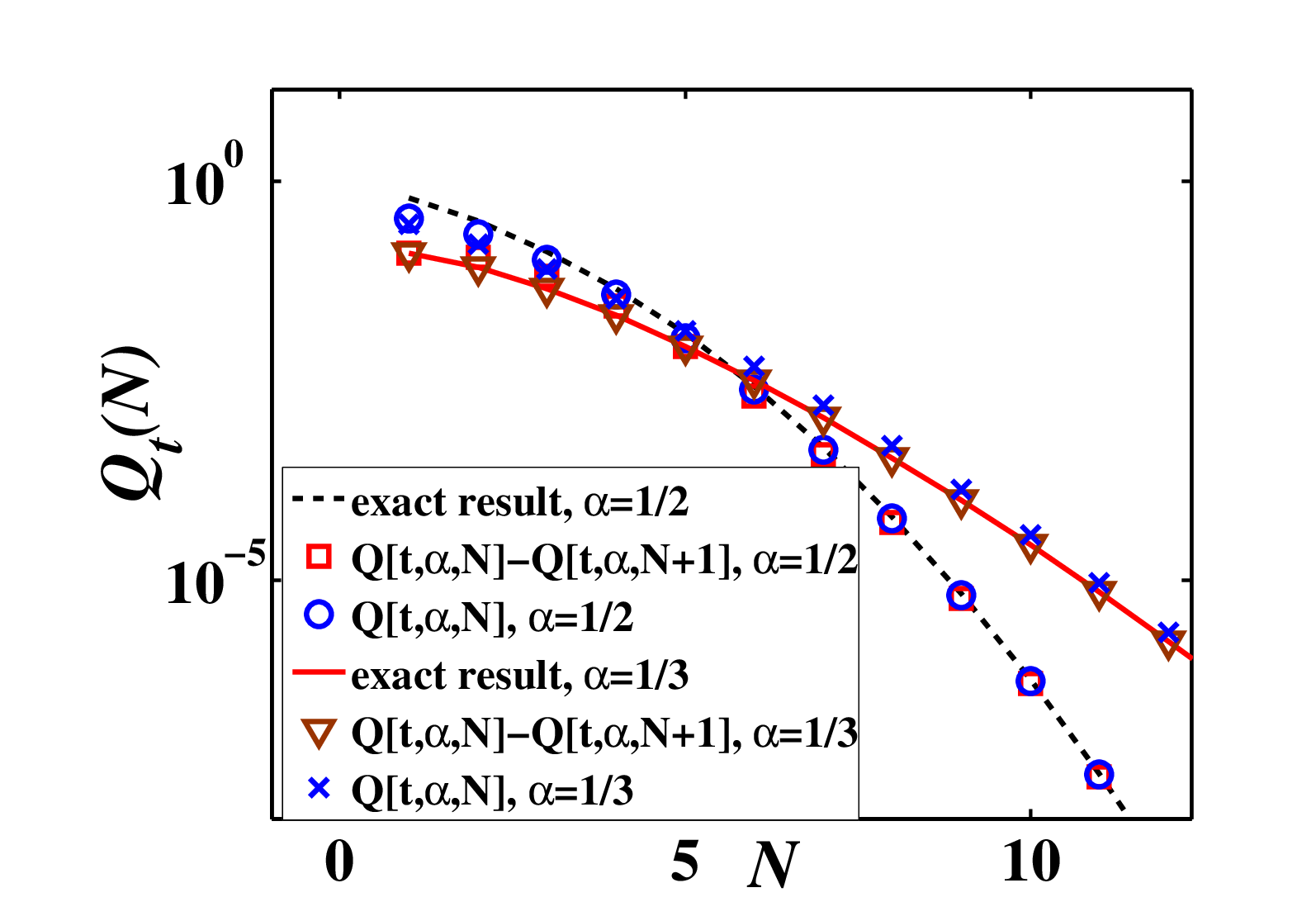}
\caption{Approximation of $Q_t(N)$ using $Q[t,\alpha,N] - Q[t,\alpha,N+1]$ and $Q[t,\alpha,N]$ for $t=2$. The symbols represent the first term ($Q[t,\alpha,N]$) and the first two terms ($Q[t,\alpha,N] - Q[t,\alpha,N+1]$) on the right-hand side of Eq.~\eqref{saddleMehodsQtn}, obtained using the saddle point approximation as the theoretical prediction. The exact results from Eq.~\eqref{eq:levyexactform}, plotted as lines, are provided for comparison.}
\label{QtnTwoTerm}
\end{figure}

First, we find the exact form of $Q_t(N)$.
Since ${\cal L}^{-1} \left\{\exp\left(-s^{\alpha}\right)\right\} = \ell_\alpha (t)$, then ${\cal L}^{-1} \left\{\exp\left(-Ns^{\alpha}\right)\right\} = \ell_\alpha (t/N^{1/\alpha})/N^{1/\alpha}$ and $Q\left[t,\alpha,N\right]=\int_0^t\ell_{\alpha}(t'/N^{1/\alpha})/N^{1/\alpha}\,dt'$. Therefore, Eq.~\eqref{saddleMehodsQtn} yields
\begin{equation}
    \label{eq:levyexactform}
    Q_t(N)  =\int^{\frac{t}{N^{1/\alpha}}}_{\frac{t}{(N+1)^{1/\alpha}}}\ell_\alpha(y)dy,
\end{equation}
which is exact for any $t$ and $N$.
In the limit of fixed $t$ and large $N$ we can use the method described in previous sections that yield for $Q\left[t,\alpha,N\right]$ in the large $N/t$ limit
\begin{widetext}
\begin{equation}\label{goodSaddlePont}
\begin{split}
&Q[t,\alpha,N]\sim\frac{\exp \left(-\alpha^{\frac{\alpha}{1-\alpha}}(1-\alpha)\left(\frac{N}{t^\alpha}\right)^{\frac{1}{1-\alpha}}-\frac{1}{1-\alpha}\ln(\frac{N}{t^\alpha})-\ln(\alpha^{\frac{1}{1-\alpha}}t)\right)}{\sqrt{2 \pi \left| \left(\frac{t}{N \alpha }\right)^{-\frac{2}{\alpha -1}} \left(1-N (\alpha -1) \left(\left(\frac{t}{N \alpha }\right)^{\frac{1}{\alpha -1}}\right)^{\alpha } \alpha \right)\right| }},
\end{split}
\end{equation}
\end{widetext}
where the saddle point is achieved at $s^{*}=(\alpha N/t)^{\frac{1}{1-\alpha}}$.
The decay to $0$ of $Q[t,\alpha,N]$ with $N$ is extremely fast and is determined by
\begin{equation}\label{sdhhsh101se}
Q[t,\alpha,N]\sim \exp\left(-(1-\alpha)N\left(\frac{\alpha N}{t}\right)^{\frac{\alpha}{1-\alpha}}\right).
\end{equation}
In Fig.~\ref{QtnTwoTerm}, we compare the exact behavior of $Q_t(N)$ (using Eq.~\eqref{eq:levyexactform}) and the approximations utilizing Eq.~\eqref{goodSaddlePont} and Eq.~\eqref{saddleMehodsQtn}. From Fig.~\ref{QtnTwoTerm}, it is clear that almost no difference exists when approximating $Q_t(N)$ by taking into account only the first term on the right-hand side of Eq.~\eqref{saddleMehodsQtn} or both terms together.

\begin{figure}[htb]
\centering
\includegraphics[width=1.0\linewidth]{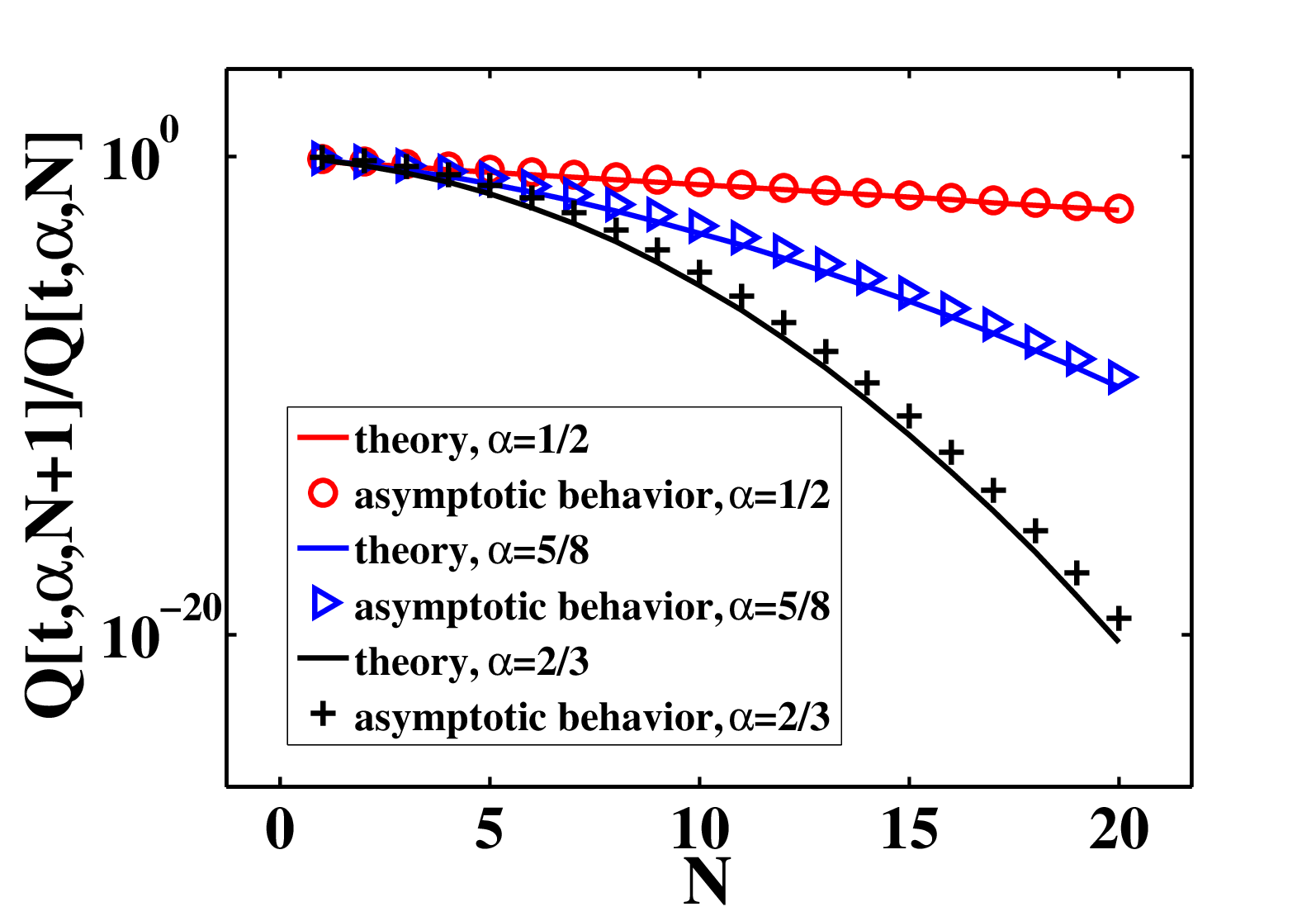}
\caption{Plot of ratio $Q[t,\alpha,N+1]/Q[t,\alpha,N]$ for various $\alpha$. The lines describe $Q[t,\alpha,N+1]/Q[t,\alpha,N]$, where $Q[t,\alpha,N]$ is obtained from Eq.~\eqref{goodSaddlePont} and the corresponding asymptotic behavior is Eq.~\eqref{calratio}. As $\alpha$ increases, the ratio $Q[t,\alpha,N+1]/Q[t,\alpha,N]$ tends to zero rapidly.  Here, we choose $t=2$.}
\label{QtnRatio}
\end{figure}

As noted above, the ratio $Q\left[t,\alpha,N+1\right]/Q\left[t,\alpha,N\right]$ is the correction we omitted in previous sections. From the comparison in Fig.~\ref{QtnTwoTerm} it is already clear that the neglected term is indeed negligible.
Nevertheless, we can explicitly inspect the behavior of the neglected term by exploring the behavior of the ratio $Q\left[t,\alpha,N+1\right]/Q\left[t,\alpha,N\right]$. From Eq.~\eqref{goodSaddlePont} we obtain in the limit of large $N/t$
\begin{equation}\label{calratio}
\begin{split}
\frac{Q[t,\alpha,N+1]}{Q[t,\alpha,N]}&\sim \exp\Big(-(1-\alpha)\left(\frac{t}{\alpha}\right)^{\frac{\alpha}{\alpha-1}}\\
&~~~~~~~\times(N^{\frac{1}{1-\alpha}}-(N+1)^{\frac{1}{1-\alpha}})\Big)\\
&\sim \exp\left(-\left(\frac{\alpha N}{t}\right)^{\frac{\alpha}{1-\alpha}}\right),
\end{split}
\end{equation}
that exhibits an extremely fast decay to $0$ with $N$ as also presented in Fig.~\ref{QtnRatio}.
It thus becomes clear why it is valid to use $Q\left[t,\alpha,N\right]$ as the approximation for $Q_t(N)$ in the large $N/t$ limit.

\section{Equilibrium Renewal Process}\label{EquilRenewal}

In Sec. \ref{model}, we have discussed the case when a physical clock started immediately at the beginning of the process, i.e., an ordinary renewal process. We further introduce the aging renewal process \cite{Monthus1996Models,Struick1978Physical,Schutz1997Single,Barkai2003Aging,Barkai2003Agingin,Schulz2014Aging,Marcin2017Aging,Boettcher2018Aging}, i.e., the process starts at $t=-t_a$, before the observation of the process starts (at
time $t=0$).
Generally, when sojourn times have an infinite mean, the observables of the process, such as the PDF of $N$, exhibit aging effects \cite{burov2010aging,Schulz2014Aging}. Here, we focus on the equilibrium renewal process \cite{Wang2019Transport},  i.e., $t_a\to\infty$. The equilibrium renewal process is also known as the stationary renewal process.
Mathematically, the only distinction between a non-equilibrium renewal process and an equilibrium one is that for the equilibrium renewal process, the first sojourn time follows
\begin{equation}\label{ssxxs}
h(\tau)=\frac{\int_\tau^\infty \phi(y)dy}{\langle\tau\rangle},
\end{equation}
instead of $\phi(\tau)$ \cite{Feller1968introduction}, where $\langle\tau\rangle$ is the average of sojourn times $\langle\tau\rangle=\int_0^\infty \tau\phi(\tau)d\tau$. Note that non-divergence of the average sojourn time $\langle \tau \rangle$ is part of the definition of the equilibrium renewal process.
The other observed sojourn periods, such as $\tau_2,\tau_3,\cdots,\tau_{n}$, are distributed according to $\phi(\tau)$, similar to the normal renewal process.
The probability of observing $N$ renewals of an equilibrium process during time $t$, $Q_{t}^{\rm eq}(N)$, differs from the probability $Q_t(N)$
 for the normal renewal process that is provided by Eq.~\eqref{AT19SEC2EQ1202}.
 Replacement of the PDF of the first sojourn time by $h(\tau)$ and the fact that ${\cal L}\left\{h(\tau)\right\}=\left(1-\widehat{\phi}(s)\right)/s\langle \tau \rangle$ leads to the following expression in Laplace space \cite{Schulz2014Aging}
\begin{equation}\label{io101}
\widehat{Q}_{s}^{\rm eq}(N)=\frac{1-\widehat{\phi}(s)}{s\langle\tau\rangle}\widehat{\phi}(s)^{N-1}\frac{1-\widehat{\phi}(s)}{s}
\end{equation}
with $N\geq1$.
When $N=0$, $Q_{t}^{\rm eq}(0)$ is the probability that the first sojourn event will outlast $t$, i.e.,  $Q_{t}^{\rm eq}(0)=\int_t^\infty \int_z^\infty \phi(y)/\langle\tau\rangle dydz$.
A critical property of the equilibrium process is that the average number of renewals follows $t/\langle\tau\rangle$ for any observation time $t$.
The behavior of $Q_{t}^{\rm eq}(N)$ in the $t\to \infty$ limit was developed for various sojourn time PDFs, as discussed in \cite{Godreche2001Statistics,Schulz2014Aging}.
Here, we focus on the large $N$ limit for fixed observation time $t$.
Though the distinction between the equilibrium and non-equilibrium renewal processes lies solely in the first sojourn time, not all observables behave the same for these two models.
For example, when the sojourn time has a finite mean but an infinite variance, rare fluctuations of the longest time intervals are different from normal renewal process \cite{Wang2019Transport}.
Therefore, it is essential to explore the effect of equilibration when considering the $N/t\to \infty$ limit.

Similarly to the case of the non-equilibrium renewal process,  the limit of large $N$ for $Q_t^{\rm eq}(N)$ is determined by the $s\to \infty$ limit of Eq.~\eqref{io101}.
Therefore, since
$\widehat{\phi}(s)\to 0$ when $t \to 0$, the leading order of Eq.~\eqref{io101} is determined by the lowest power of $\widehat{\phi}(s)$ and we can use the approximation
\begin{equation}\label{io102}
\widehat{Q}_{s}^{\rm eq}(N)\sim\frac{\widehat{\phi}(s)^{N-1}}{s^2\langle\tau\rangle}.
\end{equation}
Equation \eqref{io102} is
 valid for the case of large $N$ and fixed time $t$ in the limit when $N/t\to\infty$.
 Based on Eq.~\eqref{io102}, below we will consider two types of sojourn time PDFs.
 Namely, when $\phi(\tau)$ can be presented in terms of a power-series near $\tau\to 0$ and when a cutoff time is present.

\subsection{The case of power-series expansion of $\phi(\tau)$}\label{equillb:power}

\begin{figure}[htb]
 \centering
 \includegraphics[width=0.5\textwidth]{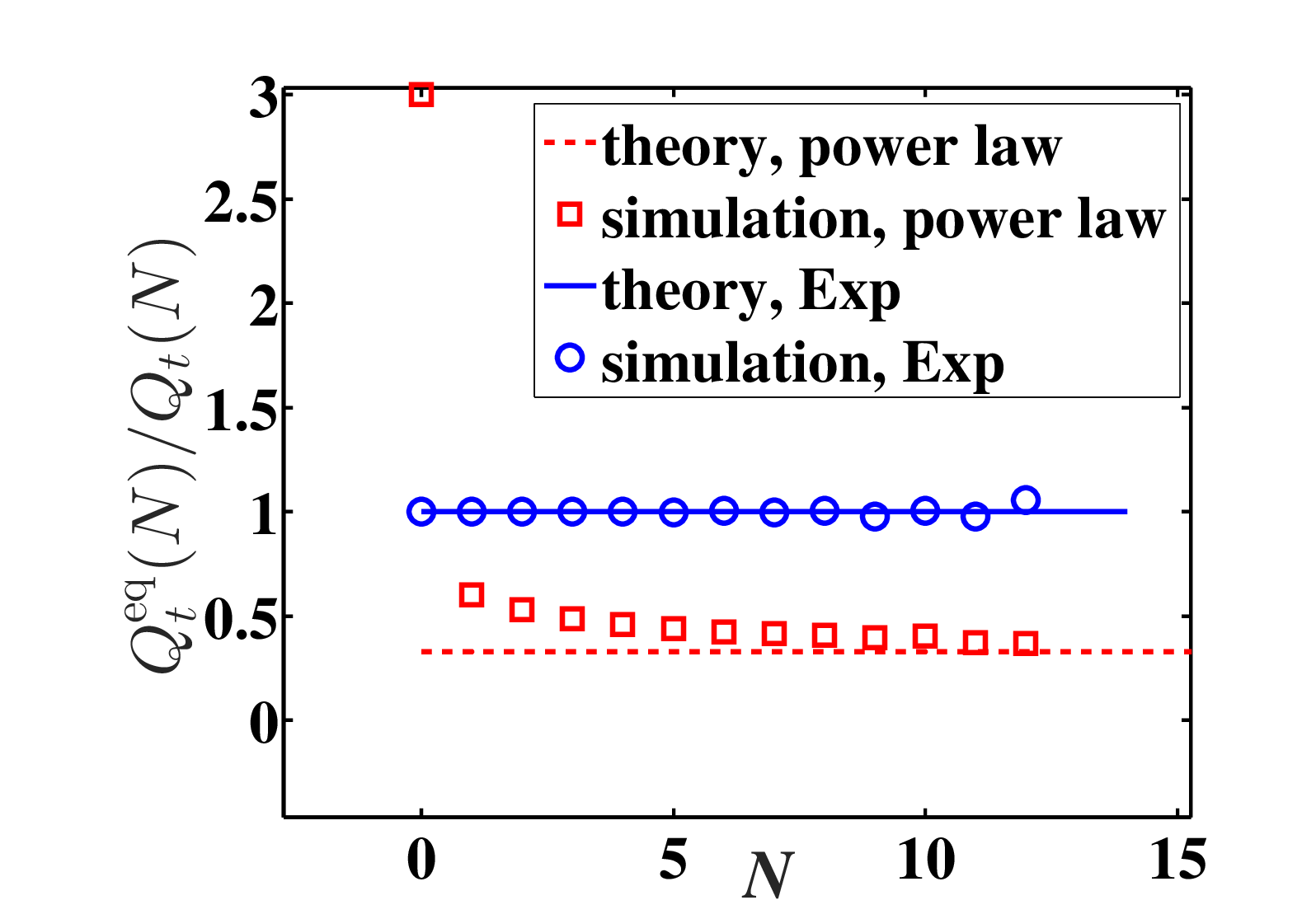}\\
 \caption{Theory and simulations of the ratio $Q_{t}^{\rm eq}(N)/Q_{t}(N)$ versus $N$ with $t=2$. Theoretical predictions for power-law distribution, i.e., $\phi(\tau)=\gamma/(1+\tau)^{1+\gamma}$ with $\gamma=3/2$, and exponential distribution $\phi(\tau)=\exp(-\tau)$  are represented by lines given in Eqs.~\eqref{ratioQtnasw} and \eqref{ratioQtnasws}, respectively.  Symbols are the simulations generated from realizations of particles.
}\label{RatioQtnForTwoTerms}
\end{figure}

Our first case involves a category of probability distributions, which can be represented by the power-series in the limit $\tau\to 0$, see Eq.~\eqref{phitausMALL}.
However, as mentioned above, the equilibrium renewal process must have a finite mean sojourn time.
Substitution of Eq.~\eqref{sdi101} into Eq.~\eqref{io102} yields
\begin{equation}
\label{eqqtnEqLap}
\widehat{Q}_{s}^{\rm eq}(N)\sim \frac{[C_\alpha \Gamma(1+\alpha)]^{N-1}}{s^{2+(1+\alpha)(N-1)}\langle\tau\rangle}\left(1+\frac{C_\beta\Gamma(1+\beta)}{C_\alpha\Gamma(1+\alpha)}s^{\alpha-\beta}\right)^{N-1}.
\end{equation}
The inverse Laplace transform of Eq.~\eqref{eqqtnEqLap}, while using the saddle point approximation for the large $N$ limit, produces
\begin{equation}\label{sjde0101ass}
\begin{split}
Q_{t}^{\rm eq}(N)&\sim \frac{t}{\langle\tau\rangle}\frac{[(C_\alpha \Gamma(1+\alpha)^{\frac{1}{1+\alpha}})t]^{(1+\alpha)(N-1)}}{\Gamma(2+(1+\alpha)(N-1))} \\
&\times\exp\left((N-1)^{1+\alpha-\beta}\frac{C_\beta\Gamma(1+\beta)t^{\beta-\alpha}}{C_\alpha \Gamma(1+\alpha)(1+\alpha)^{\beta-\alpha}}\right).
\end{split}
\end{equation}
Equation~\eqref{sjde0101ass} is the analog of Eq.~\eqref{StasExp} for the equilibrium renewal process.
Instead of analyzing numerous sojourn time PDFs to check the validity of Eq.~\eqref{sjde0101ass}, we focus on the ratio of $Q_{t}^{\rm eq}(N)$ to $Q_{t}(N)$. Based on Eqs.~\eqref{1seq201} and \eqref{sjde0101ass}, we obtain that
\begin{equation}\label{sfk101}
\frac{Q_{t}^{\rm eq}(N)}{Q_{t}(N)}\sim \frac{1}{\langle\tau\rangle C_\alpha \Gamma(1+\alpha) t^\alpha}\frac{\Gamma((1+\alpha)N+1)}{\Gamma(1-\alpha+(1+\alpha)N)}.
\end{equation}
Using Stirling's approximation,   Eq.~\eqref{sfk101} reduces to
\begin{equation}\label{sfk101addx}
\frac{Q_{t}^{\rm eq}(N)}{Q_{t}(N)}\sim \frac{(1+\alpha)^{\alpha}}{\langle\tau\rangle C_\alpha \Gamma(1+\alpha) }\left(\frac{N}{t}\right)^\alpha.
\end{equation}
In the particular case of $\alpha=0$,  Eq.~\eqref{sfk101addx} reduces to
\begin{equation}\label{ratioQtn}
\frac{Q_{t}^{\rm eq}(N)}{Q_{t}(N)}\sim \frac{1}{\langle\tau\rangle C_0}.
\end{equation}
Namely, when $\phi(\tau)$ converges to a constant $C_0$ when $\tau\to 0$, the ratio of $Q_{t}^{\rm eq}(N)$ to $Q_{t}(N)$ in the large $N$ limit is converging to a constant determined by the average sojourn time $\langle\tau\rangle$ and the first term of expansion of sojourn times distribution.
In Fig.~\ref{RatioQtnForTwoTerms} we verify the result in Eq.~\eqref{ratioQtn} by exploring numerically the behavior of the ratio $Q_{t}^{\rm eq}(N)\big/Q_{t}(N)$ as a function of $N$.
In our simulations, we use two different forms of $\phi(\tau)$, i.e., power law distribution $\phi(\tau)=\gamma/(1+\tau)^{1+\gamma}$ with $\gamma>1$  and exponential distribution $\phi(\tau)=\lambda\exp(-\lambda \tau)$. For the power-law form of $\phi(\tau)$,
$C_0=\gamma$ and Eq.~\eqref{ratioQtn} reduces to
\begin{equation}\label{ratioQtnasw}
\frac{Q_{t}^{\rm eq}(N)}{Q_{t}(N)}\sim \frac{\gamma-1}{\gamma}
\end{equation}
in the large $N$ limit.
As shown in  Fig.~\ref{RatioQtnForTwoTerms}, the convergence to the value in Eq.~\eqref{ratioQtnasw} holds for sufficiently large $N$, that in this case is around $N\ge 4$.
When $\phi(\tau)=\lambda\exp(-\lambda \tau)$, the sojourn PDF converges to $C_0=\lambda$ when $\tau\to 0$, and $\langle\tau\rangle=1/\lambda$. Therefore, for this case, Eq.~\eqref{ratioQtn} yields
\begin{equation}\label{ratioQtnasws}
\frac{Q_{t}^{\rm eq}(N)}{Q_{t}(N)}\sim 1.
\end{equation}
Equation~\eqref{ratioQtnasws} is obvious in the sense that the exponential distribution is memoryless or ageless, meaning that the equilibrium renewal process is identical to the non-equilibrium one. Thus, as stated by Eq.~\eqref{ratioQtnasws} for exponential sojourn times, $Q_{t}^{\rm eq}(N)$ is identical to $Q_{t}(N)$, not only for large $N$ but any $N$ as is verified in Fig.~\ref{RatioQtnForTwoTerms}.

\subsection{The case of $\phi(\tau)$ with a cutoff}\label{equillib:cutoff}

We now deal with the scenario when  the power-series expansion of sojourn time PDF follows Eq.~\eqref{cutoff101}, while $\langle \tau \rangle$ is finite.
According to Eqs.~\eqref{cutoff101} and \eqref{io102}, in Laplace space, $Q_{t}^{\rm eq}(N)$ follows
\begin{widetext}
\begin{equation}\label{cutoffEq}
\widehat{Q}_{s}^{\rm eq}(N)\sim \exp(-s\tau_0N)
\frac{[C_\alpha \Gamma(1+\alpha)]^{N-1}}{s^{2+(1+\alpha)(N-1)}\langle\tau\rangle}\left(1+\frac{C_\beta\Gamma(1+\beta)}{C_\alpha\Gamma(1+\alpha)}s^{\alpha-\beta}\right)^{N-1}.
\end{equation}
\end{widetext}
One method to derive the inverse Laplace transform of Eq.~\eqref{cutoffEq} is through the application of the saddle point technique. However, we notice that Eq.~\eqref{cutoffEq} differs from Eq.~\eqref{eqqtnEqLap} by a multiplicative term $\exp(-s\tau_0N)$ and therefore the shift property of the inverse Laplace transforms yields
\begin{widetext}
\begin{equation}\label{sdesd}
Q_{t}^{\rm eq}(N)\sim \frac{t-\tau_0(N-1)}{\langle\tau\rangle}\frac{[(C_\alpha \Gamma(1+\alpha)^{\frac{1}{1+\alpha}})(t-\tau_0(N-1))]^{(1+\alpha)(N-1)}}{\Gamma(2+(1+\alpha)(N-1))} \exp\left(\frac{C_\beta\Gamma(1+\beta)(t-\tau_0(N-1))^{\beta-\alpha}}{C_\alpha \Gamma(1+\alpha)(1+\alpha)^{\beta-\alpha}/(N-1)^{1+\alpha-\beta}}\right).
\end{equation}
\end{widetext}
Notice that Eq.~\eqref{sdesd} is valid when $N$ is large but $N\leq t/\tau_0+1$. Interestingly, it can be seen that the maximal value of $N$ discussed here is larger than the analog maximal value for $N$ in the non-equilibrium renewal process.
This discrepancy is due to the first sojourn time. For the equilibrium renewal process, this first sojourn time is not constrained to be greater than or equal to $\tau_0$.

From Eqs.~\eqref{sjdhs101} and \eqref{sdesd}, we obtain that the ratio $Q_{t}^{\rm eq}(N)\big/Q_{t}(N)$ follows
\begin{equation}\label{sfk101a}
\begin{split}
\frac{Q_{t}^{\rm eq}(N)}{Q_{t}(N)}&\sim \frac{(\frac{t-\tau_0(N-1)}{t-\tau_0N})^{N(1+\alpha)}}{\langle\tau\rangle[C_\alpha \Gamma(1+\alpha)](t-\tau_0 (N-1))^\alpha}\\
&~~~\times\frac{\Gamma((1+\alpha)N+1)}{\Gamma(1-\alpha+(1+\alpha)N)}.
\end{split}
\end{equation}
in the large $N$ limit.
Applying Stirling's approximation for the Gamma function in Eq.~\eqref{sfk101a} yields
\begin{equation}\label{sfk101addx23}
\begin{split}
\frac{Q_{t}^{\rm eq}(N)}{Q_{t}(N)}&\sim \frac{\exp(\alpha\ln((1+\alpha)N))}{\langle\tau\rangle C_\alpha \Gamma(1+\alpha) (t-\tau_0(N-1))^\alpha}\\
&~~~~~~~\times\left(\frac{t-\tau_0(N-1)}{t-\tau_0N}\right)^{N(1+\alpha)}.
\end{split}
\end{equation}
When $\alpha=0$, Eq.~\eqref{sfk101addx23} reduces to
\begin{equation}\label{cutoffequilibriumProcess}
\frac{Q_{t}^{\rm eq}(N)}{Q_{t}(N)}\sim \frac{1}{\langle\tau\rangle C_0}\left(\frac{t-\tau_0(N-1)}{t-\tau_0N}\right)^{N}.
\end{equation}
In particular, when $\tau_0=0$, i.e., no cut-off is present,  Eq.~\eqref{cutoffequilibriumProcess} is similar to  Eq.~\eqref{ratioQtn}.

\section{Conclusion and discussion}

We have analyzed the probability distribution of the number of renewals for both non-equilibrium and equilibrium renewal processes.
Our focus was on the behavior of rare event probabilities in the large $N/t$ limit at fixed time $t$.
In this regime, the key feature of the sojourn time PDF $\phi(\tau)$ is its asymptotic form as $\tau \to 0$.
Extending beyond prior studies that considered only the expansion in~\eqref{StasEliPlr}, we derived results for generalized power-series expansions, cutoffs, and non-analytical forms of $\phi(\tau)$.
Our approach primarily utilized the saddle point approximation and the correspondence between large-$s$ Laplace properties and short-time behavior.

In all our results, such as Eq.~\eqref{shssheq101}, Eq.~\eqref{sjdhs101}, and Eq.~\eqref{eq:nonalyticLarge}, we observe that the decay of the probability $Q_t(N)$ for large $N$ is strongly influenced by the short-time properties of $\phi(\tau)$.
These differences in the decay behavior of $Q_t(N)$ emphasize the potential of developing analytical tools to probe underlying dynamics.
When experimental data from the observation or tracking of physical, chemical, or biological processes yield histograms of a large number of events, our findings can be used to extract and interpret the short-time characteristics of $\phi(\tau)$ that govern the observed transport.
Furthermore, any experimental system will inherently have a minimal time scale below which no process can occur.
The results in Sec.~\ref{model:cutoff}
 and Sec.~\ref{equillib:cutoff} pave the way for theoretical exploration of transport properties in systems with such a minimal time scale.
Importantly, the form of $Q_t(N)$, as derived in Eq.~\eqref{sjdhs101}, vanishes beyond a certain maximal $N_{\rm max}$.
This suggests that the existence of a minimal time scale imposes bounds on the universality of Laplace tails.
Specifically, beyond the exponential decay observed in the positional PDF, additional forms of decay may be shaped by the presence of this minimal time scale in the system.
Such findings will provide new insights into deviations from universal behavior in various experimental settings.

The results in Sec.~\ref{EquilRenewal} demonstrate the behavior of $Q_t^{\rm eq}(N)$ for the equilibrium renewal process in the large $N$ limit.
While the Laplace tails were derived for the standard renewal process, it is uncertain whether these conditions apply to most experimental systems where such tails have been observed.
In many experiments, the observational clock starts after the system has evolved for some arbitrary time, making the equilibrium probability $Q_t^{\rm eq}(N)$ more relevant.
As shown in Sec.~\ref{equillb:power}
, for certain parameter sets, $Q_t(N)$ and $Q_t^{\rm eq}(N)$ exhibit similar behavior, suggesting comparable transport properties for rare events.
However, when $\phi(\tau)$ does not approach a constant as $\tau \to 0$, the ratio $Q_t^{\rm eq}(N)/Q_t(N)$ grows as $\left(N/t\right)^{\alpha}$ for large $N$ (see Eq.~\eqref{sfk101addx}), indicating significantly different transport PDF decays between the two cases.

In this work, we have focused exclusively on renewal processes, but the behavior of the probability distribution for the number of events is also a crucial quantity when the renewal assumption does not hold.
Future research should explore the behavior of this distribution in systems with correlations, such as those induced by quenched disorder~\cite{rinn2001hopping,shafir2024disorder}, or in cases involving extreme statistics, like maximal or minimal observables~\cite{mori2020distribution,singh2023universal,majumdar2024statistics}.
Exploring these factors in the large number of events limit will enhance the theoretical framework for complex systems beyond renewal dynamics.

\section*{Acknowledgments}
WW is supported by the National Natural Science Foundation of China under Grant No. 12105243 and the Zhejiang Province Natural Science Foundation LQ$22$A$050002$. SB is supported by the
Israel Science Foundation Grant No. 2796/20.



\end{document}